\documentclass [12pt]{article}
\usepackage{lscape}                                           %
\usepackage[centertags]{amsmath}
\usepackage{amsfonts} \usepackage{amssymb}\usepackage{eufrak}
\usepackage{colordvi}
\usepackage{epsfig}
\usepackage{a4}
\usepackage{cite}
\usepackage{lscape}

\newcommand{\COMMENTO}[1]{}

\begin{document}
\begin{titlepage}
\rightline{DSF-INFN/1-2009}
\rightline{NORDITA-2009-6} \vskip 3.0cm
\centerline{\LARGE \bf  K{\"{a}}hler Metrics:
}\vskip .5cm
\centerline{\LARGE \bf  String vs Field Theoretical Approach}
\vskip 1.0cm \centerline{\bf P. Di Vecchia$^{a,b}$, A. Liccardo$^{c,d}$, R. Marotta$^d$
and F. Pezzella$^d$}
\vskip .6cm \centerline{\sl $^a$
The Niels Bohr Institute, Blegdamsvej 17, DK-2100 Copenhagen \O, Denmark}
\vskip .4cm
\centerline{\sl $^b$
Nordita, Roslagstullsbacken 23, SE-10691 Stockholm, Sweden}
\vskip .4cm \centerline{\sl
 $^c$ Dipartimento di
Scienze Fisiche, Universit\`a degli Studi ``Federico II'' di Napoli}
\centerline{\sl Complesso Universitario Monte
S. Angelo ed. 6, via Cintia,  I-80126 Napoli, Italy}
\vskip .4cm
\centerline{\sl
$^d$ Istituto Nazionale di Fisica Nucleare, Sezione di Napoli}
\centerline{ \sl Complesso Universitario Monte
S. Angelo ed. 6, via Cintia,  I-80126 Napoli, Italy}

 \vskip 1cm

\begin{abstract}
We use both a string and a field theoretical approach in order to determine respectively the K\"ahler metrics of untwisted and  twisted open strings attached to magnetized D9 branes in toroidal compactifications. 
\end{abstract}
\vfill  {\small{ Work partially supported by the European
Community's Human Potential Programme under contract
MRTN-CT-2004-005104 ``Constituents, Fundamental Forces and Symmetries
of the Universe''. }}
\end{titlepage}

\newpage

\tableofcontents       %
\vskip 1cm             %

\section{Introduction}
\label{intro}

In order to compare the predictions of string theory with experimental data it is necessary to determine the low-energy four-dimensional effective Lagrangian involving chiral matter fields and their supersymmetric partners. It is well-known that any ${\cal N}=1$ SUGRA Lagrangian in $D=4$ is characterized by three functions of the chiral multiplet: the K\"ahler potential (real) $K$, the superpotential (holomorphic) and the gauge kinetic functions  (holomorphic). In particular the kinetic terms $Z_{IJ}D_{m}\bar{\phi}^{I}D^{m}\phi^{J}$ of the complex scalars exhibit as coefficients the K\"ahler metric $Z_{IJ}$ of the manifold spanned by them. It is  
defined as the second-order derivative of $K$.
It has to be possible to compute these quantities from string theory.
Magnetized braneworld  provides a promising scenario where this program could be achieved. Indeed in this framework one can construct explicit D brane models
 which at low energy yield the gauge group of the Standard Model and four-dimensional chiral fermions.
 It can be realized by space-time filling D9 branes with a background constant magnetic field switched on along the six compact space dimensions of the ten-dimensional space-time $R_{1,3} \otimes M_{6}$ being $M_{6}$ a suitable compact manifold. 
 Within these brane configurations different kinds of massless fields are generated: the ones associated to the closed string states living in the bulk, i.e. gravitational fields and scalar moduli fields generated by the internal space compactification; the ones coming from {\em dipole} or {\em untwisted} open strings attached to a single magnetized brane or to the worldvolume of a stack of branes having the same magnetizations, providing the gauge sector and, finally, the ones relative to the {\em dycharged} or {\em twisted} open strings stretched between two (stacks of) magnetized branes with different magnetizations, corresponding to four-dimensional quarks and lepton fields. In the case of a constant magnetization, the dynamics of the open strings can be completely and analytically determined and the open strings can be exactly quantized. This means that from the computation of string amplitudes one could in principle determine the low-energy four-dimensional effective Lagrangian. It will depend on the microscopic data specifying the D brane configurations and the geometry of the compact space. From it one can read the K\"ahler metrics with their full dependence on moduli \cite{lust, bertolini}. Alternatively, if not interested in string corrections of those quantities, one can start from the action of ${\cal N}=1$ SYM in D=10, that describes the low-energy dynamics of the open strings attached to D9 branes, perform on it the Kaluza-Klein reduction from ten to four dimensions and derive the low-energy four-dimensional effective action for the massless excitations \cite{cremades, CMQ, DLMP}. In this paper we use the string approach for computing the K\"ahler metrics of the untwisted strings and the above mentioned field theoretical one for getting the K\"ahler metrics of the twisted strings.


\section{Magnetized branes in string theory: spectrum and field theory limit.}
\label{magnetized}

In this section we analyze the spectrum of twisted open strings and  perform on it the field theory limit ($\alpha' \rightarrow 0$) showing that one gets the same spectrum obtained from the field theory approach. More specifically, we analyze
a system with two stacks of D9 branes whose worldvolume is partially wrapped on the factorized torus $T^{2(1)} \otimes T^{2(2)} \otimes T^{2(3)}$.  A constant magnetic field with field strength $F_{a}^{(r)}\equiv F^{(r)}_{a} dx^{2r+2} \wedge dx^{2r+3}$ ($r=1,2,3$) is turned-on on the $r$-th torus around which a stack of D branes, denoted with $a$,  is wrapped $n_{a}$ times. Since on a torus $T^{2(r)}$ the first Chern class is an integer $I_{a}^{(r)} \in Z$ one has:
\begin{equation}
\frac{1}{2\pi} \int_{T^{2(r)}} \mbox{Tr} (F_{a}^{(r)}) = I^{(r)} _{a} \in Z \rightarrow 2 \pi \alpha' n_{a} F^{(r)}_{a} = I^{(r)}_{a}\equiv T_{2}^{(r)} \mbox{tan} \pi \nu^{(r)}_{a}
\label{1}
\end{equation}
where the last equation defines the {\em magnetization} of the stack $a$ of branes. $T_{2}^{(r)}$ is the K\"ahler modulus given by the physical volume of the $r$-th torus in units of string length: $
T_{2}^{(r)}\equiv V_{T^{2(r)}}/(2 \pi \sqrt{\alpha'})^2.$

\subsection{Bosonic spectrum}

In the NS sector, after defining the number operators along the non-compact directions:
\begin{eqnarray}
N^{\alpha} = \sum_{n=1}^\infty\alpha_{-n} \alpha_n \hspace{1.5cm}; \hspace{1.5cm}
N^{\psi} =\sum_{n=1/2}^{\infty} n \psi_{-n}\psi_n
\label{NXNpsi}
\end{eqnarray}
and those along the compact directions:
\begin{eqnarray}
N^{a} =
\sum_{r=1}^{3} \left[
\sum_{n=0}^{\infty} (n + \nu^{(r)} ) a^{(r)\dagger}_{n+\nu^{(r)}}  a_{n+\nu^{(r)}} +
\sum_{n=1}^{\infty} (n - \nu^{(r)} ) \bar{a}^{(r) \dagger}_{n-\nu^{(r)}} \bar{a}_{n-\nu^{(r)}}
\right]
\label{NcalZ}
\end{eqnarray}
\begin{eqnarray}
N^{\Psi}= \sum_{r=1}^3\! \,\,\,
\!\Big[   \sum_{n=1/2}^\infty (n+\nu^{(r)})
\Psi^{\,\dagger \,(r)}_{n+\nu^{(r)}}\Psi^{\, (r)}_{n+\nu^{(r)}}
\!+     \sum_{n=1/2}^\infty (n-\nu^{(r)})
{\overline \Psi}^{\,\dagger \,(r)}_{n-\nu_r}{\overline \Psi}^{\, (r)}_{n-\nu^{(r)}}
\Big],
\label{NPsi}
\end{eqnarray}
we obtain the mass spectrum of the  open string states  stretched between two stacks, labelled with $a$ and $b$,  of differently magnetized D9 branes:

\begin{eqnarray}
M^2 = \frac{1}{\alpha'} \left[N^{\alpha} + N^{\psi} + N^{a} + N^{\Psi} - \frac{1}{2}
+ \frac{1}{2} \sum_{r=1}^{3} \nu^{(r)}     \right] 
\label{M2}
\end{eqnarray}
where $\nu^{(r)} =\nu^{(r)}_{a} - \nu^{(r)}_{b}$ is the difference of the magnetizations of the two stacks.

The lowest bosonic states in the NS sector are the following:
\begin{eqnarray}
{\bar{\Psi}}^{\dagger (r)}_{1/2 - \nu^{(r)}} |0 >~~r=1,2,3  \hspace{1.5cm} \mbox{and}  \hspace{1.5cm}
{\bar{\Psi}}^{\dagger (1)}_{1/2 - \nu^{(1)}} {\bar{\Psi}}^{\dagger (2)}_{1/2 - \nu^{(2)}}
{\bar{\Psi}}^{\dagger (3)}_{1/2 - \nu^{(3)}}|0 >.
\label{senzama}
\end{eqnarray}
Their mass is respectively: 
\begin{eqnarray}
M^{2}_{r} = \frac{1}{\alpha'} \left[ \frac{1}{2} \sum_{s=1}^{3} \nu^{(s)} - \nu^{(r)} \right] \hspace{1cm}; \hspace{1.5cm}
M^{2}=  \frac{1}{\alpha'}\left[ 1 -  \frac{1}{2} \sum_{s=1}^{3} \nu^{(s)} \right].
\label{M2b}
\end{eqnarray}
One of these states becomes massless when any of the following conditions
is satisfied:
\begin{eqnarray}
 \frac{1}{2} \sum_{s=1}^{3} \nu^{(s)} - \nu^{(r)}=0 \hspace{.5cm};\hspace{.5cm}
 \sum_{s=1}^{3} \nu^{(s)} =2
\label{susyco}
\end{eqnarray}
corresponding to have ${\cal{N}}=1$ supersymmetry.
We may consider also the states
\begin{eqnarray}
{{\Psi}}^{\dagger (r)}_{1/2 + \nu^{(r)}} |0 >r=1,2,3 \hspace{1.5cm}\mbox{and} \hspace{1.5cm}
{{\Psi}}^{\dagger (1)}_{1/2 + \nu^{(1)}} {{\Psi}}^{\dagger (2)}_{1/2 + \nu^{(2)}}
{{\Psi}}^{\dagger (3)}_{1/2 + \nu^{(3)}}|0 >
\label{senzama2}
\end{eqnarray}
with a mass respectively given by:
\begin{eqnarray}
M^{2}_{r} = \frac{1}{\alpha'} \left[ \frac{1}{2} \sum_{s=1}^{3} \nu^{(s)} + \nu^{(r)}\right]
\hspace{.5cm};\hspace{.5cm}
M^{2}=  \frac{1}{\alpha'}\left[ 1 + \frac{3}{2} \sum_{s=1}^{3} \nu^{(s)} \right].
\label{M2bc}
\end{eqnarray}

\subsection{Field theory limit}

Let us start to analyze the field theory limit performed as $\alpha' \rightarrow 0$ while keeping the volume  $V_{T^{2(r)}}$ fixed, which implies necessarily small values of the magnetization $\nu^{(r)}$
given by 
\begin{eqnarray}
\pi \nu^{(r)}= |I^{(r)}|/T_{2}^{(r)}
\label{nusmall}
\end{eqnarray}
that can be obtained from Eq. (\ref{1}). In the NS sector, one might think that, because of the factor $1/\alpha'$ in Eqs. (\ref{M2b}) and (\ref{M2bc}) only
the massless states survive in the field theory limit. But this is not quite true.
Because of  Eq. (\ref{nusmall}) the states that survive in the field
theory limit are only those that have a mass squared linear in $\nu^{(r)}$
without any additional numerical constant. It seems therefore that, in the
field theory limit, one has to keep the first states in Eqs. (\ref{M2b}) and (\ref{M2bc}),
while the last two ones get an infinite mass. Actually, it is not so because one can have any excitation of the bosonic oscillators
with $n=0$. This means that the mass formula for the surviving states reads as:
\begin{eqnarray}
(M_{\pm \,\,n}^{(r)})^2  & = &  \lim_{\alpha' \rightarrow 0}   \frac{1}{\alpha'} \left[
\sum_{s=1}^{3} \nu_s \left(
 a^{\dagger (s)}_{\nu_s}  a^{(s)}_{\nu_r}   + \frac{1}{2} \right)
 \pm  \nu_r \right]  \nonumber \\
 & = & \lim_{\alpha' \rightarrow 0}   \frac{1}{2 \alpha'} \left[
\sum_{s=1}^{3} \nu_s \left( 2 N_s    + {1} \right)
 \pm  2 \nu_r \right]
\label{MpmN}
\end{eqnarray}
where $N_s$ is the bosonic number operator.

The next step consists in deriving Eq. (\ref{MpmN}) within a purely field theoretical approach \cite{cremades,DLMP} as we are going to illustrate.

\section{Magnetized branes: field theoretical approach}

The starting point is the action of ${\cal N}=1$ SYM in D=10 with gauge group $U(N)$ which can also be regarded as the worldvolume action of a stack of $N$ space-time filling D9 branes.
\begin{equation}
S= \frac{1}{g^{2}} \int d^{10}X \sqrt{-G} \,\,\mbox{Tr} \left[ -\frac{1}{4} F_{MN}F^{MN} + \frac{1}{2} \bar{\lambda} \Gamma^{M}D_{M} \lambda \right] .
\label{action}
\end{equation}
The main point in this description is to decompose the gauge field $A_{M}$ and the gaugino $\lambda$ as follows:
\begin{equation}
A_{M}=B_{M}+W_{M}=B_{M}^{a}U_{a}+W_{M}^{ab}e_{ab}\,\,\,\,\,; \,\,\,\,\, \lambda=\chi+\psi=\chi^{a}U_{a}+ \psi^{ab} e_{ab}
\end{equation}
where the generators $U_{a}$ are in the Cartan subalgebra differently from the remaining ones denoted by $e_{ab}$ ($a \neq b$). By suitably choosing the background values of the gauge fields, one could interpret $S$ as a theory describing stacks of D9 branes with different magnetizations. This can be realized first by decomposing $B_{M}^{a}(X)$ and $W_{M}(X)$ along the four non-compact dimensions $x^{\mu}$ ($\mu=0,\dots,3$)  and the six
compact ones $x^{i}$ ($i=4, \dots, 10$), $X^{M}=(x^{\mu},x^{i})$. One can give arbitrary vacuum expectation values to $B_{k}^{a}$ and $W_{k}$ as follows:
\begin{eqnarray}
B^{a}_{k} (x^{\mu},x^{i}) & = & < B^{a}_{k} > (x^{i}) + \delta B^{a}_{k} (x^{\mu}, x^{i}) \nonumber \\
W_{k}^{ab}(x^{\mu}, x^{i}) & = & 0+ \partial W_{k}^{ab} (x^{\mu}, x^{i})  \,\,\, \equiv \,\,\,\Phi_{k}^{ab} (x^{\mu},x^{i} ).
\label{PHI}
\end{eqnarray}
The presence of different background values along the Cartan subalgebra breaks the original gauge group $U(N)$ in $U(1)^{N}$  generating $N$ stacks, each 
consisting of one brane, having  different magnetizations. In this picture the field $\Phi_{k}^{ab}$ defined in (\ref{PHI}) describes twisted open strings. The spectrum of the Kaluza-Klein excitations is obtained by expanding the ten-dimensional fields on a basis of eigenstates of the corresponding internal wave function:
\begin{equation}
\Phi^{ab}_{i} (x^{\mu},x^{i}) = \sum_{n} \varphi_{i,n}^{ab} (x^{\mu})  \phi_{i,n}^{ab}(x^{i})
\end{equation}
and solving the equation
\begin{equation}
-\tilde{D}_{k} \tilde{D}^{k} (\phi^{ab})_{i,n} =  \sum_{r=1}^{3} \frac{2 \pi | I^{(r)}|}{T_{2}^{(r)}} (2 N^{(r)}+ 1 ) \phi^{ab}_{i,n} = m^{2}_{n} \phi^{ab}_{i,n} (x^{i})
\end{equation}
obtaining  the eingenfuctions written in Ref. \cite{cremades, DLMP}. One finds two towers of Kaluza-Klein states for each torus with masses given by:
\begin{eqnarray}
(M_{n, r}^{\pm})^{2} = m_{n}^{2} \pm  \frac{4 \pi I^{(r)}}{(2 \pi \sqrt{\alpha'})^2
{T}_{2}^{(r)} } =
\frac{1}{(2 \pi \sqrt{\alpha'})^2} \left[\sum_{s=1}^{3}
\frac{2 \pi |I^{(s)}| }{T_{2}^{(s)}}
\left( 2N_s +1 \right) \pm
\frac{4 \pi I^{(r)}}{T_{2}^{(r)}} \right]. 
\label{KKmass}
\end{eqnarray}
This formula shows that one
can have a massless state  for $N_{s}=0$ only if,  using Eq. (\ref{nusmall}), the first condition in Eq. (\ref{susyco}) is satisfied.
In this case one  keeps ${\cal{N}}=1$ supersymmetry.

 Eq. (\ref{KKmass}) has to be compared with the spectrum  previously obtained in the string approach, i.e. with Eq. (\ref{MpmN}).  In fact, inserting Eq.  (\ref{nusmall}) in the latter, one can see that the factors $\alpha'$
cancel and that the field theoretical expression in Eq. (\ref{KKmass}) is recovered.

In conclusion, the bosonic sector of the twisted strings reproduces, in the zero slope limit, the one
obtained by using the field theoretical method described above.

One can  show that the same happens also in the fermionic sector.

\section{String computation of the K\"ahler metric for the adjoint}

Our goal is to compute the K\"ahler metrics for untwisted and twisted strings with their full dependence on the moduli. We start from the former case by using the string approach. In this case the moduli are given by the parameters of the three-torus $T^{2(1)} \otimes T^{2(2)} \otimes T^{2(3)}$. In particular, each torus is parametrized by the K\"ahler and complex structure moduli, $T=T_{1}+iT_{2}$ and $U=U_{1}+iU_{2}$.  Computing a K\"ahler metric in string theory means computing a string two-point function. This is not a straightforward task since all fields are canonically normalized in the string basis so it would be impossible to read the K\"ahler metric as a function of all moduli.

In the literature a method has been proposed to overcome this problem \cite{lust, bertolini}. The recipe consists in calculating the disk amplitude $A(M, \phi, \bar{\phi} ) \sim <V_{M}V_{\phi}V_{\bar{\phi}}>  $  involving the vertex operators of two scalar fields (on the boundary) and of a closed string modulus and to get a differential equation for the K\"ahler metric that, once integrated, fixes it.  This procedure has allowed to explicitly get the dependence on the modulus $T$ but not, even in the simplest case of the untwisted scalars,  the one on the modulus $U$.
This latter has been instead obtained by factorizing a four-point amplitude \cite{lust}. Furhermore, using the same procedure, in Ref. \cite{bertolini} the four-dimensional effective action has been obtained in terms of the ten-dimensional real fields.
In the following we would like to show that it is possible to catch the explicit dependence on the modulus $U$  by computing still a suitable three-point amplitude.  Using the notation of Ref. \cite{bertolini}, we can write the open string vertex for the scalar 
correspondent to the $r$th torus $T^{2(r)}$ as follows:
\begin{eqnarray}
V_{A^{(r)}}^{(0)} = i N_{A} A_{m}^{(r)} \frac{(1+R_{0}^{(r)})^{m}_{\,\,n}}{2} 
\left( \partial_{z} X^{n}_{r} (z) + i \sqrt{2 \pi \alpha'} (k \cdot \psi) \Psi^{n}_{r}) 
\right) e^{ \sqrt{2 \pi \alpha'} i k \cdot X(z)}
\label{Vert}
\end{eqnarray}
where the indices  $r=1,2,3$ label the three tori and $m,n=1,2$ run over  the two ``curved" coordinates of each  torus $T^{2}$,
$R_{0}^{(r)}$ stands for the reflection matrix $R_{\sigma}^{(r)}$ \cite{pesando} defined on the endpoint 
$\sigma=0$ of the string. In Eq. ({\ref{Vert}) $\psi$ and $\Psi$ denote the world-sheet fermions  along the non-compact and compact directions respectively.
It results to be more convenient to work in the basis diagonalizing the matrix $R_r \equiv R^{-1(r)}_{\pi} R_{0}^{(r)}$, because the modes in the solution for the string coordinates are given in terms of the eigenvalues of this matrix. The relation between the two basis is given by:
\begin{eqnarray}
X^{a}_{r} \equiv\left(\begin{array}{c}
{\bar{Z}}_r\\
{Z}_r
 \end{array}\right)=({\cal{E}}_{r})^{a}_{\,\,m} X^{m}_{r} 
 =\sqrt{\frac{T_{2}^{(r)}}{2 U_{2}^{(r)}}}\left(\begin{array}{c}
                                            X^{1}_{r}+U_r X^{2}_{r}\\
                                            X^{1}_{r}+{\bar{U}}_r X^{2}_{r}
                                            \end{array}\right)
\label{XAXM}
\end{eqnarray}
and similarly for $\Psi^m$,
where ${\cal E}_r$ is the matrix which diagonalizes $R_r$ and  also $R_{0}^{(r)}$ on $T^2$:
\begin{eqnarray}
({\cal{ E}}_r )^a_{\,\,m}(R_{r})^m_{\,\,n}({\cal E}^{-1}_r)^{n}_{\,\,b}=
({\cal R}_r )^a_{\,\,b}\equiv \left( \begin{array}{cc} e^{2i \pi \nu_r }  & 0 \\
                                 0  &  e^{-2i \pi \nu_r } \end{array} \right) .
\label{diag4}
\end{eqnarray}
In the following, for the sake of simplicity, we restrict ourselves to consider the case  
$\nu_{r}=0$.

Written in the new basis,  the vertex (\ref{Vert}) 
becomes: 
\begin{eqnarray}
V_{A_r}  =  V_{\Phi_{r}}+V_{\bar{\Phi}_{r}} 
\label{VAr}
\end{eqnarray}
where:
\begin{eqnarray}
V_{\Phi_r }&=&iN_A\Phi_r (\partial_z {\bar{Z}}_r  +i \sqrt{2\pi\alpha'}
k\cdot \psi {\bar{\Psi}}_{r})e^{\sqrt{2\pi\alpha'} ik\cdot X}\nonumber\\
V_{\bar{\Phi}_r}&=&iN_A\bar{\Phi}_r(\partial_z Z_{r} +i \sqrt{2\pi\alpha'}
k\cdot \psi \Psi_{r})e^{\sqrt{2\pi\alpha'} ik\cdot X} \,\,.
\label{14}
\end{eqnarray}
The relation between the two real fields $A_1 , A_2$ and the complex one 
$\Phi$ is given by: 
\begin{eqnarray} 
( {\bar{\Phi}} ,  \Phi)  & \equiv  & A_m  \left( \frac{1 +R_0}{2} \right)^{m}_{\,\,n}   
( {\cal{E}}^{-1} )^{n}_{\,\,a} \nonumber \\
& = &  \frac{i}{ \sqrt{2 T_{2}^{(r)} U_{2}^{(r)} }} 
( (A_{1}^{(r)} {\bar{U}}_r - A_{2}^{(r)}) ,  - ( A_{1}^{(r)} {{U}}_r - A_{2}^{(r)}))\,\,.
\label{conre}
 \end{eqnarray}
The string vertex for the modulus $U$, in the complex basis, is given by \cite{lust}:
\begin{eqnarray}
V_{U^{(r)}}^{(-1,\,-1)}=i\frac{N_G}{U_2^{(r)}}\Psi_r (z) {\tilde{\Psi}}_r 
(\bar{z}) e^{\sqrt{2 \pi \alpha'} ik\cdot X(z,\,\bar{z})}
e^{-\phi(z)}\,e^{-\bar{\phi}(\bar{z})} .
\label{VG1}
\end{eqnarray}
Using the above vertices one can show that the only non-vanishing amplitude is:
\begin{eqnarray}
{\cal A}_{U_{r}\bar{\Phi}_r\bar{\Phi}_r}&=& C_{0}  \int\frac{d^2 z_3dz_1dz_2}{dV_{abc}}\langle
V_{\bar{\Phi}_r}(z_1,\,k_1)\,V_{\bar{\Phi}_r}(z_2,\,k_2)\,
V_{U^{(i)}}(z_3,\,\bar{z}_3,\,k_3,\,k_4)
\rangle .
\label{nonvamp}
\end{eqnarray}
Taking into account the correlators \cite{lust,bertolini, pesando}:
\begin{equation}
 \langle \partial_{z} Z_r (z) \partial_{w} {\bar{Z}}_s (w) \rangle = - 
 \frac{\delta_{rs}}{(z-w)^2} \hspace{1cm} ; \hspace{1cm}
 \langle \Psi_r (z) {\bar{\Psi}}_{s} (w) \rangle =
 - \frac{\delta_{rs}}{z-w} 
 \end{equation}
 \begin{equation}
\langle \partial_{z} Z_r (z) \partial_{w} {{Z}_s} (w) \rangle =
\langle \partial_{z} {\bar{Z}}_r (z) \partial_{w} {\bar{Z}}_s(w) \rangle = 
\langle \Psi_r (z) \Psi_s (w) \rangle = 
\langle {\bar{\Psi}}_r (z) {\bar{\Psi}}_s (w) \rangle = 0\,, 
\end{equation}
after a straighforward calculation, Eq. (\ref{nonvamp}) results to be:
\begin{eqnarray}
{\cal A}_{U^{(r)}\bar{\Phi}_r\bar{\Phi}_r}=\frac{8}{U_{2}^{(r)}}\,
\sqrt{\pi} 2^{-2\pi\alpha't-3}\, (N_G\,N_A^2\,C_0)(\bar{\Phi}_r)^2 \pi \alpha'\,t\frac{\Gamma(\frac{1}{2}-\pi\alpha't)}{\Gamma(1-\pi\alpha' t)}
\end{eqnarray}
with $t=- k_{1} \cdot k_{2}$ \cite{lust}. 
Using the following expression for the normalization factors:
\begin{eqnarray}
N_A= i \sqrt{\pi\alpha'}g_{YM}  \hspace{.5cm};\hspace{.5cm}C_0=\frac{4}{g_{YM}^2(2\pi\alpha')^2} \hspace{.5cm}; \hspace{.5cm}
N_G = \frac{i}{2 \pi}
\label{nor94}
\end{eqnarray}
and performing the field theory limit
one finds:
\begin{eqnarray}
{\cal A}_{U^{(r)}\bar{\Phi}_r\bar{\Phi}_r}=  k_1 \cdot k_2 (\bar{\Phi}_r)^{2} \frac{i}{2 U_2^{(r)}} \,\,.
\label{AAUfi67}
\end{eqnarray}
The previous amplitude must be equal to the derivative of the scalar action in the supergravity basis:
\begin{eqnarray}
{\cal A}_{U\bar{\Phi} \bar{\Phi}}=K(\Phi,\bar{\Phi})^{-1} \partial_U {\cal L} \hspace{1cm};
\hspace{1cm} {\cal{L}} = k_1 \cdot k_2 \, \Phi  \, {\bar{\Phi}}
\end{eqnarray}
where we have omitted the index $r$ and which, under the hypothesis that the fields are holomorphic functions of the moduli just like the superpotential is,  yields the following differential equation involving the K\"ahler metric $K(\Phi,\bar{\Phi})$:
\begin{eqnarray}
\!\!\!\!\!\!\Phi(U) \partial_U\log K(\Phi,\bar{\Phi})+ \partial_U \Phi(U) =\frac{i}{2U_2}\bar{\Phi}(\bar{U})
\label{diff}
\end{eqnarray}
and its complex conjugate.
By differentiating these equations with respect $\bar{U}$ and $U$ respectively, we get
the following identity:
\begin{eqnarray}
\bar{\Phi}(\bar{U})\left[\bar{\Phi}(\bar{U})+(U-\bar{U}) \partial_{\bar{U}}\bar{\Phi}(\bar{U})\right]=
\Phi({U})\left[\Phi({U})-(U -\bar{U})\partial_{U}\Phi({U})\right]
\end{eqnarray}
solved by requiring that:
\begin{eqnarray}
\Phi({U})-(U -\bar{U})\partial_{U}\Phi({U})= \bar{\Phi}(\bar{U})\label{difs}
\end{eqnarray}
together with its complex conjugate. 

Using the Laurent expansion $
\Phi(U)=\sum_{n\in Z}a_n(x) U^n
$ in Eq. (\ref{difs}) one gets: 
\begin{eqnarray}
\sum_{n\in Z}\left[ (1-n)a_n(x) U^n + a_n(x) nU^{n-1} \bar{U}-\bar{a}_n(x)\bar{U}^n\right]=0.
\end{eqnarray}
The previous identity can be satisfied only if $n=0,1$ and $a_n=\bar{a}_n$ 
and plugging it in Eq. (\ref{diff}) yields:
\begin{eqnarray}
(a_0(x)+a_1(x)\,U)\partial_U\log K+a_1(x)= -\frac{a_0 +a_1 \bar{U}}{U-\bar{U}}.
\end{eqnarray}
Being the fields $a_0(x)$ and $a_1(x)$ independent, one  can write:
\begin{eqnarray}
U\partial_U\log K +1 =-\frac{\bar{U}}{U-\bar {U}} \Longrightarrow \partial_{U} \log K = - \frac{1}{U- \bar{U} }
\end{eqnarray}
whose solution is given by  $K \sim \frac{1}{U_2}.$

In this way, we have obtained the dependence of the K\"ahler metric on the modulus 
$U$ for the untwisted strings. 

\section{Field theoretical approach for the computation of the K\"ahler metrics}.

The K\"ahler metric for twisted ${\cal N}=1$ matter fields can be computed by using the field theoretical approach described above. The main ingredient consists
in not canonically normalizing the kinetic terms but keeping the moduli dependence that naturally comes in the Kaluza-Klein reduction.
Let us consider the K\"ahler metric of the massless
scalar that we name  $\varphi$. The
K{\"{a}}hler metric $Z$ can be read from its kinetic term given by:
\begin{eqnarray}
-  \int d^4 x \sqrt{G_4} \,\,Z( m , {\bar{m}} ) \,
(D_{\mu} {\bar{\varphi}}  (x) )
( D^{\mu} \varphi  (x) )
\label{Kaeme}
\end{eqnarray}
written in the Einstein frame.  After performing the integral over the torus one gets:
\begin{eqnarray}
Z & = & 
\frac{{\rm e}^{  \phi_{4}}}{2} N_{\varphi}^{2} \prod_{r=1}^{3}  \left[
\left(\frac{1}{2  U_{2}^{(r)}} \right)^{1/2}
\left( \frac{T_{2}^{(r)} }{ | I_r | } \right)^{1/2} \right] \nonumber \\ & = &
\frac{N_{\varphi}^{2}}{2s_{2}^{1/4}}
\prod_{r=1}^{3} \left[ \frac{ 1}
{(2 u_{2}^{(r)})^{1/2}  (t_{2}^{(r)})^{1/4}} \left( \frac{T_{2}^{(r)} }{ | I_r | } \right)^{1/2} \right] .
\label{Zkahler0}
\end{eqnarray}
In Eq. (\ref{Zkahler0}) we have introduced the moduli in the supergravity basis related to the ones in the string basis by the following relations:
\begin{equation}
t_{2}^{(r)} \equiv e^{- \phi_{10}} T_{2}^{(r)} \,\,\,\,\, ;  \,\,\,\,\,  u_{2}^{(r)}=U_{2}^{(r)} \,\,\,\,\, ;  \,\,\,\,\, s_{2}= e^{-\phi_{10}} \prod_{r=1}^{3} T_{2}^{(r)}
\end{equation}
with $\phi_{10}$ and $\phi_{4}$ denoting, respectively, the ten- and the four-dimensional dilaton.  
Furthermore, $N_{\varphi}$ is a normalization function that is fixed by requiring the holomorphicity of the superpotential. It can be obtained by evaluating the Yukawa couplings  for the chiral multiplet. In fact they have the following structure:
\begin{equation}
Y_{ijk} = e^{K/2}W_{ijk}
\end{equation}
where $K$ is the K\"ahler potential and $W_{ijk}$ is the superpotential which is a holomorphic function of the fields.
The explicit expression of the Yukawa couplings has been calculated in Ref. \cite{cremades, DLMP}.  It contains a product of the three normalization factors $N_{\varphi}^{ab}, N_{\psi}^{bc}, N_{\psi}^{ca}$ of the
four-dimensional fields involved in the Yukawa coupling and these can be chosen in such a way to eliminate the non-holomorphic dependence on the magnetizations.
In so doing we get for the chiral scalars:
\begin{eqnarray}
Z^{chiral}_{ab}  = \frac{1}{2 s_{2}^{1/4}}
\prod_{r=1}^{3} \left[ \frac{ 1}{(2 u_{2}^{(r)})^{1/2}  (t_{2}^{(r)})^{1/4}} \right]
\left( \frac{\nu^{(1)}_{ab}  }{\pi  \nu^{(2)}_{ab} \nu^{(3)}_{ab} } \right)^{1/2} .
\label{Zkaehlerf}
\end{eqnarray}

The dependence of the K\"ahler metric on the
magnetizations has been computed by means of a pure string calculation in
Refs.~\cite{lust,bertolini}. This dependence has been confirmed by
instanton calculations that, however, have also provided
an additional dependence on the moduli and possibly also on the 
magnetizations arriving at the following K\"{a}hler metric \cite{tedeschi1, tedeschi2, istant}:
\begin{eqnarray}
Z^{chiral}_{ab} \sim \frac{ \Gamma(1-\nu_{ab}^{(1)})  \Gamma(\nu_{ab}^{(2)} ) \Gamma(\nu_{ab}^{(3)} )}{\Gamma(\nu_{ab}^{(1)})  \Gamma(1-\nu_{ab}^{(2)}) \Gamma(1-\nu_{ab}^{(3)}) }
 \prod_{r=1}^{3} \left[ \left( u_{2}^{(r)} \right)^{-\xi \nu_{ab}^{(r)}} \left( t_{2}^{(r)} \right)^{-\zeta \nu_{ab} ^{(r)}   } \right]
 \label{zchiral}
\end{eqnarray}
being $\xi $ and $\eta$ arbitrary parameters. Eq. (\ref{zchiral}) exactly reproduces
Eq. (\ref{Zkaehlerf})   in the field theory limit which is realized for small values
of  $\nu_{ab}^{(r)}$.

\begin{center}
{\bf Acknowledgements}
\end{center}
 F. P. thanks the organizers of the 4th RTN Workshop in Varna where he presented this work that was partially supported by the European Community's Human Potential Programme under contract MRTN-CT-2004-005104 ``Constituents, Fundamental Forces and Symmetries of the Universe".

\end{document}